\documentstyle[aps,prl]{revtex}
\begin{document}
\draft
\def\la{\langle}
\def\ra{\rangle}
\def\om{\omega}
\def\Om{\Omega}
\def\vep{\varepsilon}
\def\wh{\widehat}
\newcommand{\beq}{\begin{equation}}
\newcommand{\eeq}{\end{equation}}
\newcommand{\beqa}{\begin{eqnarray}}
\newcommand{\eeqa}{\end{eqnarray}}
\newcommand{\intf}{\int_{-\infty}^\infty}
\newcommand{\into}{\int_0^\infty}
\input epsf
\preprint{EHU-FT/0009,quant-ph/0009111}
\twocolumn
\begin{title}
{\Large \bf Model for the arrival-time distribution in
fluorescence time-of-flight experiments}
\end{title}
\author{J. G. Muga$^1$, A. D. Baute$^{1,2}$,
J. A. Damborenea$^{1,2}$ and I. L. Egusquiza$^2$}
\address{$^1$ Departamento de Qu\'\i mica-F\'\i sica, Universidad del
Pa\'\i s Vasco, Apdo. 644, 48080 Bilbao, Spain}
\address{$^2$ Fisika Teorikoaren Saila, Euskal Herriko Unibertsitatea,
644 P.K., 48080 Bilbao, Spain}

\maketitle

\begin{abstract}

An operational arrival-time distribution is defined as the
distribution of detection times of the first photons
emitted by two level atoms in resonance with a perpendicular laser
beam in a time of flight experiment.  For ultracold Cesium atoms the
simulations are in excellent agreement with the theoretical ideal
time-of-arrival distribution of Kijowski.
\end{abstract}

\pacs{PACS: 03.65.-w\hfill EHU-FT/0009,quant-ph/0009111}

In many experiments the observables are the {\it instants} when
certain events occur, or the {\it durations} of processes.  This is
also the case in experiments where the scale of the physical system
requires a quantum treatment.  Nevertheless, and, in retrospect,
rather surprisingly, considering time as an observable has been a
long standing taboo for theorists because of Pauli's argument
against the existence of a self-adjoint time operator conjugate to a
Hamiltonian bounded from below \cite{Pauli}, a condition that applies
to most systems of physical interest.

It is however quite clear nowadays, even though not yet a widespread
piece of knowledge, that ``Pauli's theorem'' is not a major stumbling
block to formalize time in quantum mechanics, first because actual
observables are in general linked to positive operator valued measures
(POVM) that do not require self-adjoint (first-moment) operators
\cite{BGL95}, and second because in fact the ``theorem'' itself does
not hold unless due care is taken of the domains of the operators
implied (see for instance the analysis in\cite{Galapon}).

While getting rid of the hindrance of Pauli's theorem is an important
step, there still remains the question of defining time observables
and/or time distributions.  One then immediately runs into the difficulty
of dealing with the ``occurrence of events'' in quantum mechanics,
since the standard formalism provides a smooth, continuous description
of the state evolution, without any explicit abruptness such as
``collapses'' or ``quantum jumps'' associated with the events.  After
all, this may not be surprising if the quantum equations and the state
are statistical and refer to ensembles; even clasically, abrupt
changes for individuals (such as a yes/no transition) may be averaged
out by statistical ensembles.  The question then is how to extract
information for individuals from the formalism, and in particular on
the instants when the events occur.  The current technical ability to
trap and manipulate single atoms has triggered theoretical approaches
to deal with a number of interesting time observables.  These
techniques make compatible the continuous and the jump descriptions,
for the ensemble and the individual respectively.  For example,
the statistics of the dark periods of a fluorescing system with a
rapidly decaying state and a metastable state have been theoretically
understood and reproduced \cite{CD}.
Our aim is to apply a similar analysis to
arrival-time experiments.

Among different time observables, the time of arrival (TOA) is one of
the simplest, and an excellent study case since ``arrival times'' are
routinely measured in particle, nuclear, atomic, and molecular physics
laboratories.  There has been much recent interest among theorists
on the arrival time, as reported in an extensive recent review
\cite{ML}.  In several works the possibility of defining ideal TOA
distributions has been denied
\cite{Allcock,YT}, or claims
have been made that the arrival time cannot be measured precisely
\cite{AOPRU}.  However, these negative conclusions have not been
shared by all.  In particular, Kijowski's TOA distribution for free
motion has been defined and characterized as an optimal ideal
distribution in the sense of satisfying in a unique manner several
classically motivated properties \cite{Kijowski74}.  Nevertheless,
there is a clear divorce between the daily routine of many
laboratories where time-of-flight experiments are performed, and the
theoretical studies on the time of arrival, which are based on the particle's
wave function without recourse to extra (apparatus) degrees of
freedom.  A number of ``toy models'' have been proposed that include
simple couplings between the particle's motion and other degrees of
freedom acting as clocks or stopwatches, but they do not incorporate
any irreversibility and are still far from realistic experimental
conditions \cite{AOPRU}.
Halliwell has provided so far the only
irreversible model for a time-of-arrival measurement based on a two
level detector where the initial excited level decays due to the
presence of the particle.  The model remains however rather abstract
and no identification is made with any specific measuring system
\cite{Halliwell98b}.

Advancing in this direction, we provide in this letter a theoretical
model to fill the gap between theory and experiment by capturing
essential ingredients of current atomic time-of-flight experimental
settings, where the arrivals are measured as detections of
fluorescence photons \cite{SGAD}.  Our model is based on a number of
approximations and idealizations, which are either standard ones, or
can be easily improved upon to make it even more accurate and
realistic. In the experiment we model, the atoms are first prepared in
a certain state, then fly a distance freely, and finally cross a
perpendicular laser beam, resonant with one of the atomic transitions.
The laser intensity is here assumed to increase sharply to a constant
value in the direction of atomic propagation to facilitate the
comparison with existing ideal TOA theories, and in particular with
Kijowski's distribution.  Note however that this ``step'' profile may
be easily modified, e.g.  by using a more realistic Gaussian shape.
The first fluorescence photon detected marks operationally the arrival
time since it is assumed that all emitted photons are detected by a
perfect counter.  The preparation is such that a single atom density
operator will adequately describe the dynamics.  Even though it is not
strictly necessary for the model setting, we shall deal in the
simulations with ultracold atoms at velocities of the order of
centimeters per second.

The atom's evolution can be modelled by a one dimensional, two
component, quantum master equation.  The rotating wave and electric
dipole approximations, and a semiclassical description of the
atom-laser coupling are assumed; the internal structure of the atom is
reduced to the two levels coupled by the laser, $|1\ra$ and $|2\ra$,
with zero detuning.  The interaction picture based on the internal
state atomic Hamiltonian $H_{int}$,
\beq
H_{int}=\frac{\omega \hbar}{2} (|2\ra\la2|-|1\ra\la 1|),
\eeq
is also used since it lets us eliminate the frequency $\omega$
and the time
explicitly from the Hamiltonian. Finally, assuming
a vacuum state background (this is plausible because of the high value of
optical frequencies), the master equation takes the form
\beqa\label{me}
\frac{d\rho}{dt}=-\frac{i}{\hbar}[H,\rho]+\frac{\gamma}{2}\{
2 \sigma^- \rho \sigma^+-\sigma^+ \sigma^- \rho -\rho \sigma^+\sigma^-\},
\eeqa
where
\beqa
H&=&p^2/2m
+\frac{\hbar}{2} \Omega(x)(\sigma^+ + \sigma^-),\label{ha}\\
\sigma^-&=&|1\ra\la 2|,\nonumber\\
\sigma^+&=&|2\ra\la 1|,
\nonumber
\eeqa
$\Omega(x)$ is the position dependent Rabi frequency
($\Theta(x)\Omega_0$ in the simulations), and $\gamma$ is the inverse
of the lifetime of the excited state due to the coupling with the bath
of modes of the background vacuum-field.

Note that the master equation in Halliwell's model is formally similar
\cite{Halliwell98b}.  There are however some important differences.
In our case the two levels correspond to the internal structure of the
particle, whereas in Halliwell's model they are assigned to the
detector.  Our atom begins in the ground state, it is excited by the
laser and decays emitting a photon, whereas Halliwell's detector
starts in an excited state and decays due to the coupling with the
particle, which is restricted to a half-line.  Halliwell's Hamiltonian
in (\ref{me}) has only diagonal kinetic terms, and lacks the
non-diagonal Rabi-frequency terms of (\ref{ha}) which are responsible
for the absorption and emission induced by the laser.  This leads in
Halliwell's model to a closed dynamical equation for the population of
the excited state which does not hold here.  Finally, while our
dissipative term acts for all $x$, since an excited atom may also
decay out of the laser field after the excitation, Halliwell's decay
term is restricted spatially and is in fact the interaction term.

The solution of the master equation (\ref{me}) may be carried
out by the techniques variously known as
``quantum trajectories'' \cite{Carmichael}, ``Monte Carlo
wave function approach'' \cite{DCM}, or ``quantum jumps''
\cite{Hegerfeldt93,PK98}. These methods do not only allow to solve
the master equation; they
also provide, associated with specific measurement
procedures, a temporal history or quantum trajectory for
the individuals where the time observables can be read directly.
For our two level system, repeated measurements
of the spontaneously emitted  photons at
time intervals $dt<\Omega_0^{-1},\gamma^{-1}$
(so that at most only one spontaneous photon is emitted in $dt$),
lead, on a coarser time scale, to a Schr\"odinger equation for
the two component state vector conditioned to no photon detection,
$|\psi_c\ra$,
\beq
\la x|\psi_c\ra=|1\ra \la x|\psi_{c,1}\ra+ |2\ra\la x|\psi_{c,2}\ra.
\eeq
This Schr\"odinger equation is
governed by an effective
Hamiltonian,
\beq\label{ef}
H_{eff}=H-i\hbar\frac{\gamma}{2}|2><2|.
\eeq
The imaginary term responsible for the decay leads to a probability of
detection of a spontaneously emitted photon in the interval
$dt$ as given by $p_c=\gamma dt P_2$, where $P_2$ is the probability of the
excited state (note the caveats on identifying emitted and detected
photons in \cite{PK98}).  To
reproduce the master equation, the ensemble is separated at every time
step into detected or undetected cases by random collapses to the
ground state according to $p_c$ \cite{PK98}.  Thus, each atom's history
is characterized by a ``trajectory'' with a record of fluorescent
photon detections at certain time steps.  Averaging over many
trajectories the master equation is recovered, but at variance with
the smooth description of the master equation, each trajectory
contains a sequence of collapses comparable to the experimental record
of time-resolved photon counts for a single atom.  For current
purposes, only the time of detection of the first photon is relevant,
so in our case the time dependent norm of the state vector in the
Schr\"odinger equation with the effective Hamiltonian $H_{eff}$
provides directly the operational arrival-time distribution $\Pi(t)$
we were seeking, as identical to the ``delay function'' \cite{CD,ZMW},
\beqa
\Pi(t)&=&-dN/dt,
\\
N&=&\sum_{j=1,2}\la \psi_{c,j}(t)|\psi_{c,j}(t)\ra,
\eeqa
without the need of a random ``Monte Carlo'' sampling.  This
simplifies greatly the numerical treatment.

In the simulations we have used the split operator method
to solve the two-component Schr\"odinger equation with the
effective Hamiltonian (\ref{ef}). For simplicity  
a minimum-uncertainty-product wave packet
for a Cesium atom is prepared out of the laser range,
with average velocities currently achieved by laser cooling,
and with its center 1.05 $\mu m$ away from the origin. Note that
it is possible to simulate longer flight paths since
the free motion of the Gaussian wave-packet is known analytically.
A closer modelling of specific experimental conditions may
require combining several minimum uncertainty
packets, for example to reproduce the spatially broad states generated
in a magneto-optical trap, or the direct propagation of initially mixed
states rather than pure states.
The velocity standard deviation, $\delta v=0.098 {\rm~cm}/{\rm s}$, is
taken in agreement with  actual experimental conditions
\cite{SSDD}.
We set $\gamma= 2\pi\times 5.3 {\rm~MHz}$ as corresponds  to the
Cesium resonance line for the transition $6^2S_{1/2}\leftrightarrow6^2P_{3/2}$
at 852 nm, with a lifetime of 30.0 ns.

\begin{figure*}[t]
\centering
\leavevmode\epsfysize=6cm \epsfbox{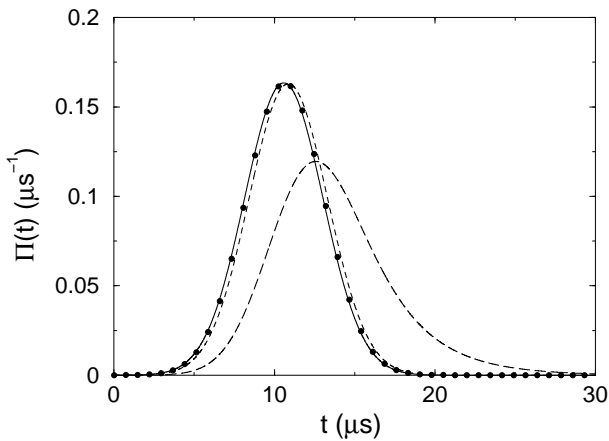}\\
\caption[image]{\label{image} The continuous line corresponds to
Kijowski's distribution for the initial wavefunction described in the
text. The longer dashed line is the operational time of arrival
distribution for $\Omega_{0}=0.099\gamma$. Shorter dashes are
associated to $\Omega_{0}=0.372\gamma$, and dots to $\Omega_{0}=1.24\gamma$.}
\end{figure*}

The figure shows Kijowski's TOA distribution for the free motion of
the initial state, with an average velocity of 10 cm/s, and the
operational (normalized) TOA distributions obtained for three
different values of $\Omega_0$ ($0.099\gamma$, $0.372\gamma$, and
$1.24\gamma$).  For the smallest value the operational distribution is
clearly delayed with respect to Kijowski's, whereas for the two larger
values the agreement is excellent.  A remarkable finding is that the
agreement is not broken by the significant fraction of atoms ($9\%$)
eventually rejected by the laser field barrier for the largest $\Omega_0$.
Essentially the same pattern with respect to the three values of
$\Omega_0$  has been also found for
larger (40 cm/s) and smaller (1 cm/s) incident velocities.
In the later case, only $20\%$ of the atoms provide a fluorescence
photon for $\Omega_0=1.24\gamma$.

This model opens up many opportunities for further insight into other
time observables, such as dwell times, or tunnelling times.  In the
case of the arrival time, it will stimulate the currently faint
interaction between experiment and theory of time observables.  It may
also help the experimentalists to select beforehand optimal Rabi
frequencies and lifetimes.  For the theorist, it is a simple tool to
compare results of different approaches with simulations of
experiments.  An important bonus of the time distribution based on the
first fluorescence photons is that it can be equally applied for free
motion or in the presence of potentials (other than those associated
to the measurement itself), so that it provides an operational
first-arrival-time distribution for which ideal (apparatus
independent) theoretical distributions are not as well studied as for
the free motion case.

\acknowledgments{We are grateful to S. Brouard for
helpful discussions.
This work has been supported
by Ministerio de Educaci\'on
y Cultura (Grants
PB97-1482 and AEN99-0315), The
University of the Basque Country (Grant UPV 063.310-EB187/98),
and the Basque Government (PI-1999-28).
A. D. Baute acknowledges an FPI fellowship by Ministerio de
Educaci\'on y Cultura.}

\end{document}